%% file: paper.tex
\newcommand{\diff}{\text{d}}

\newenvironment{psmallmatrix}
  {\left(\begin{smallmatrix}}
  {\end{smallmatrix}\right)}

\documentclass {article} 

\usepackage{amsmath,amsfonts,amssymb,amsthm,mathrsfs}
\usepackage{dsfont}
\usepackage{mathtools,float}
\usepackage{mathrsfs}
\usepackage{booktabs}
\usepackage{mathtools}
\usepackage[hidelinks]{hyperref}
\usepackage{tikzit}
\input{Pics/Style.tikzstyles}
\usetikzlibrary{shapes.geometric}
\usepackage{braket}
\renewcommand\bra[1]{{\langle{#1}|}}
\makeatletter
\renewcommand\ket[1]{%
  \@ifnextchar\bra{\k@t{#1}\!}{\k@t{#1}}%
}
\newcommand\k@t[1]{{|{#1}\rangle}}
\makeatother
\usepackage[tight]{minitoc}
\usepackage{titletoc}
\usepackage{tocloft}

\titlecontents{section}
[0pt]                                               
{}%
{\small \contentsmargin{5pt}                               
    {\bfseries \thecontentslabel}\enspace}
{\contentsmargin{0pt}\large}                        
{\small \titlerule*[.5pc]{.}\contentspage}                 
[]   
\usepackage{sectsty}

\subsubsectionfont{\normalfont\itshape}

\usepackage{accents}

\begin{document}

\title{A new 2+1 coherent spin-foam vertex for quantum gravity}
\author{J. D.\ Simão\footnote{Institute for Theoretical Physics, Friedrich-Schiller-University Jena, Helmholtzweg 4, 07743 Jena, Germany. Email: \href{j.d.simao@uni-jena.de}{j.d.simao@uni-jena.de}.}}
\date{\today}
\maketitle

\begin{abstract}
\noindent We propose an explicit spin-foam amplitude for Lorentzian gravity in three dimensions, allowing for both space- and time-like boundaries. The model is based on two main requirements: that it should be structurally similar to its well-known Euclidean analog, and that geometricity should be recovered in the semiclassical regime. To this end we introduce new coherent states for space-like boundary edges, derived from the continuous series of unitary $\mathrm{SU}(1,1)$ representations. We show that the relevant objects in the amplitude can be written in terms of the defining representation of the group, just as so happens in the Euclidean case. We derive an expression for the semiclassical amplitude at large spins, showing it relates to the Lorentzian Regge action.
\end{abstract}

\tableofcontents
\vskip-0.2em\noindent\hrulefill

\section{Introduction} 

Spin-foam models \cite{Baez:1999sr, Perez:2012wv, Kaminski:2009fm}  are a class of quantum gravity hypotheses born out of the developments of loop quantum gravity (LQG). Their general structure is composed of 1) a characterization of the set of states which describe the boundary of a given space-time region, and 2) an amplitude map assigned to those very states. Whatever the explicit form of the amplitude, all spin-foams are constructed from a discretization of the associated classical space-time; most frequently, this discretization amounts to a simplicial triangulation. 

A curious property of two of the most recent and successful Lorentzian spin-foam proposals - the BC \cite{Barrett:1999qw} and EPRL \cite{Engle:2007wy} models - is that, in constructing the quantum amplitude map from the discretized classical theory in 4 dimensions, one is forced to make a choice of causal character for the sub-simplices in the triangulation. That is, the various triangles and tetrahedra must be assumed to be individually either space-like, time-like or null. Although the first incarnation of the BC and EPRL models had assumed for simplicity all such cells to be space-like, later work has generalized this state of affairs to combinations of space- and time-like regions \cite{Perez:2000ep, Conrady:2010kc,Jercher:2022mky}. 

In the absence of empirically verifiable predictions at the current state of development of the field, making contact with the classical theory has become a research priority. Much work has been devoted to the semiclassical limit of spin-foams, and recent years have seen progress in clarifying the behavior of the EPRL model (as well as the Conrady-Hnybida (CH) extension to time-like regions) in that regime.  It is a general result \cite{Barrett:2009mw,Kaminski:2017eew,Liu:2018gfc,Simao:2021qno} for this class of models that, under certain assumptions, the amplitude associated to a 4-simplex asymptotically relates to the cosine of the Regge action (a discretized version of the Einstein-Hilbert action). That the quantum theory satisfies to some extent its classical expectations evidently reinforces the reasonableness of the approach. 

Unfortunately, this pleasing circumstance seems to face some obstacles in the case of a 4-simplex with time-like triangles. Not only is the explicit formula for the amplitude particularly convoluted when compared to the remaining cases \cite{Liu:2018gfc}, it also happens that the dominant configurations in the amplitude are not isolated: effectively, a fixed choice of boundary does not yield a unique geometry in the semiclassical regime, as would otherwise be expected \cite{Simao:2021qno}. While formulas for the semiclassical amplitude of the EPRL model have been derived for all other types of boundary \cite{Barrett:2009mw,Kaminski:2017eew}, an expression for time-like triangles remains unknown; indeed, we would claim that the semiclassical amplitude is yet to be fully understood, and hence that the emergence of the Regge action is not yet guaranteed. Moreover, the finiteness of the time-like model has not been proven, while all other amplitudes were shown to be finite \cite{Engle:2008ev, Kaminski:2010qb}.  

It is in the above manner that the need for a reformulation of the time-like amplitude presents itself. As a step towards a well-defined and complete amplitude for Lorentzian gravity in 4 dimensions, we propose to explore the problem in the more accessible context of $2+1$ dimensions. Doing so will allow us to further clarify the difficulties of the 4-dimensional theory, and to develop insights regarding their solution. 

{\centering \noindent\rule{2cm}{0.3pt} \\~\\}

\noindent In this paper we propose a spin-foam amplitude for Lorentzian 3-dimensional gravity with both space- and time-like boundaries. The vertex amplitude is derived from the first-order tetrad formulation of general relativity with gauge group $\mathrm{SU}(1,1)$ (the double cover of the rotation group of $\mathbb{R}^ {1,2}$) \cite{Freidel:2000uq}. The main novelty resides in the introduction of a new set of boundary states for space-like triangle edges (the lower-dimensional analogs of the problematic time-like triangles in 4 dimensions\footnote{The geometric interpretation of boundary states hinges on the semiclassical analysis of the vertex amplitude. Minkowski's theorem allows one to characterize tetrahedra by their \textit{normal} face vectors. Triangles, on the other hand, are characterized by a closure relation between vectors \textit{parallel} to the edges.}), constructed from the continuous series of unitary irreducible  $\mathrm{SU}(1,1)$ representations. These states are coherent, following the insights of \cite{Livine:2007vk}, but differ from the original proposal of Conrady and Hnybida \cite{Conrady:2010kc}. We show that the amplitude constructed from such states - once appropriately regularized - scales with the power of the spin of the relevant representation space, in full analogy to what happens with the Euclidean model (see e.g. \cite{Perez:2012wv}). The problem of non-isolated critical points, alluded to in the previous paragraphs, is surmounted by an explicit inclusion of a Gaussian gluing constraint in the vertex amplitude. The classical limit of the model is derived, and the Lorentzian Regge action \cite{Asante:2021phx} is recovered.

\section{The 3d Euclidean model as a paradigm}

In order to clarify the requirements we will make for the Lorentzian model, it is worthwhile to review the structure of the usual spin-foam amplitude for Riemannian 3-dimensional gravity. 

One starts by taking a simplicial complex of tetrahedral cells, assigning an amplitude to each tetrahedron - the vertex amplitude.  In the coherent state formulation \cite{Livine:2007vk}, to every edge of every triangle in the boundary of a tetrahedron correspond states of the form
\begin{equation}
  \label{cohst}
  \!\scalebox{0.65}{\input{Pics/cstate.tikz}}:= \ket{j,n}:=D^j(n) \ket {j, j}\,, \quad n\in \mathrm{SU}(2)\,,
  \end{equation} 
where $D^j$ is a unitary irreducible representation of $\mathrm{SU}(2)$ with spin $j$, acting on the associated support Hilbert space $\mathcal{H}^j$. The vector $\ket {j, j} \in \mathcal{H}^ j$ is a maximal-weight eigenstate of both the Casimir and the usual $L^3$ generator ($L^3:=\sigma_3/2$ in the defining representation). Under a choice of spins $j_{ab}$ and group elements $n_{ab}$ at each edge $ab$ shared between triangles $a$ and $b$, the vertex amplitude reads
\begin{equation}
  \scalebox{0.35}{ \input{Pics/vertexsu2.tikz}}=\int_{\mathrm{SU}(2)}\prod_{a=1}^4 \diff g_a\, \prod_{a\neq b} \braket{j_{ab},n_{ab}|D^{j_{ab}}(g_a)^\dagger D^{j_{ab}}(g_b)|j_{ab},n_{ba} }\,.
\end{equation}
As is evident from its definition, the vertex amplitude amounts to a convolution of $\mathcal{H}^j$ inner-product pairings between coherent states, there being one such pairing for each pair of neighboring triangles.

The unitary representations of $\mathrm{SU}(2)$ satisfy the Clebsch-Gordan isomorphism
\begin{equation}
\label{su2tensor}
 I:\;  \bigotimes^{2j}_{i=1}{\mathcal{H}^{\frac{1}{2}}} \simeq V \oplus \mathcal{H}^j\,,
\end{equation}
where $\mathcal{H}^j$ is the supporting Hilbert space of the fundamental representation, and $V$ is a reducible representation. Since $D^{{\otimes 2j} }(L_3)\ket{\frac{1}{2} \, \frac{1}{2}}^{\otimes 2j}=j$, it must be that $I\circ\ket{\frac{1}{2} \, \frac{1}{2}}^{\otimes 2j}=\ket{j,j}$. Thus each of the $\mathrm{SU}(2)$ inner products appearing in the amplitude can be written in the defining representation as
\begin{align*}
  \braket{j_{ab},n_{ab}|D^{j_{ab}}(g_a)^\dagger D^{j_{ab}}(g_b)|j_{ab},n_{ba} }&= \braket{1/2,n_{ab}|g_a^\dagger g_b|1/2,n_{ba} }^{2j_{ab}}\\
  &= \bra{+}n_{ab}^\dagger g_a^\dagger g_b n_{ba} \ket{+}^{2j_{ab}}\,,
\end{align*}
where $\ket{+}:=\begin{psmallmatrix}1 \\0\end{psmallmatrix}$ (and $\ket{-}:=\begin{psmallmatrix}0 \\1\end{psmallmatrix}$ for the future). Equivalently, setting $\ket{z_{ab}}:=n_{ab} \ket{+} \in \mathbb{C}^2$, 
\begin{equation}
  \label{scale}
  \braket{j_{ab},n_{ab}|D^{j_{ab}}(g_a)^\dagger D^{j_{ab}}(g_b)|j_{ab},n_{ba} }= \braket{z_{ab}| g_a^\dagger g_b| z_{ba}}^{2j_{ab}}\,.
\end{equation}
The vertex amplitude can then alternatively be characterized by a pairing of Weyl spinors $\ket{z_{ab}}$ and $ \ket{z_{ba}}$ under the canonical $\mathbb{C}^2$ inner product, rotated by $\mathrm{SU}(2)$ matrices and scaled with the power of $j_{ab}$. 

We would like to argue that equation \eqref{scale} reflects a critical property of the Euclidean amplitude: that the boundary data can be formulated in terms of spinor pairings (see \cite{Wieland:2014vta} for an Euclidean spin-foam action solely in terms of spinors), and that it scales with the representation label. Indeed, previous works on the ``twisted geometries'' framework of LQG \cite{Freidel:2010aq, Freidel:2010bw, Livine:2011gp} (see also \cite{Rennert:2016rfp}) have established that the classical phase space $T^* \mathrm{SU}(2)$ of a spin network link can be modeled on $\mathbb{C}^2 \oplus \mathbb{C}^2$ with an appropriate symplectic structure, and moreover that such normalized spinors can be mapped to unit 3-vectors as
\begin{equation}
  v_z:= \braket{z|\sigma^i|z}\hat{e}_i \in S^ 2 \subset \mathbb{R}^3 \,, \quad \ket{z}:=\begin{pmatrix}
z_1 \\ z_2  \end{pmatrix} \in \mathbb{C}^2\,,
\end{equation}
$\sigma^i$ being the standard Pauli matrices. It is moreover well-known that the representation label $j_{ab}$ (related to the area spectrum in LQG) parametrizes the interface between the quantum and classical regimes of spin-foam models \cite{Barrett:2009mw,Livine:2007vk}, the latter being attained at arbitrarily large spins. Equation \eqref{scale} therefore implements a pairing which involves the geometrical quantities one expects in the classical limit, weighted by a parameter of classicality. 

In constructing a 3d Lorentzian model we will assume the above properties to be desirable, with the necessary adaptations to the objects of a Lorentzian theory. Instead of $\mathrm{SU}(2)$ we will consider the universal cover of the identity component of the rotation group in Minkowski $\mathbb{R}^{1,2}$ space, i.e. $\mathrm{SU}(1,1)$. Instead of the Weyl spinors which characterize the classical phase space $T^*\mathrm{SU}(2)$, spinors adapted to $T^*\mathrm{SU}(1,1)$  ought to be recovered. As it was recently shown in \cite{Simao:2024zmv}, these spinors are of two types: 1) Weyl spinors $\ket{z}=\begin{psmallmatrix}z_1 \\z_2 \end{psmallmatrix}$, geometrically corresponding to the two-sheeted hyperboloid $H^\pm$ (note $\varsigma:=(\sigma_3, i\sigma_2, -i\sigma_1)$)
\begin{equation}
  \begin{gathered}
    \label{2sheet}
    \ket{z}\;\, \mathrm{s.t.} \; \braket{z|\sigma_3|z}^2=1  \; \mapsto \; v_z=\pm \braket{z|\sigma_3 \varsigma^i |z}\hat{e}_i \in H^\pm\,, \\
  H^\pm =\{(t,x,y) \in \mathbb{R}^{1,2} \;| \; t^2-x^2+y^2=1\,, \; \pm t >0  \}\,;
  \end{gathered}
  \end{equation}
and 2) Majorana spinor pairs\footnote{Objects reminiscent of Majorana spinors have been employed in the literature for describing the continuous series of $\mathrm{SL}(2,\mathbb{R})$ representations. See e.g. \cite{Wieland:2020ogk, Wieland:2024dop}, where such representations are realized in terms of two-component real spinors.} $\left\{\ket{z^i}=\begin{psmallmatrix} z_i \\ \overline{z}_i\end{psmallmatrix}\right\}_{i=1,2}$, assigned to the one-sheeted space-like hyperboloid
\begin{equation}
\label{1sheet}
  \begin{gathered}
    \ket{z^i} \;\, \mathrm{s.t.} \; \braket{z^2|\sigma_3|z^1}^2=-1  \; \mapsto \; v_z^i =\braket{z^2|\sigma_3 \varsigma^i|z^1}\hat{e}_i \in H^{\mathrm{sl}}\,, \\
  H^{\mathrm{sl}} =\{(t,x,y) \in \mathbb{R}^{1,2} \;| \; t^2-x^2+y^2=-1 \}\,.
  \end{gathered}
\end{equation}

The following two sections are dedicated to the construction of $\mathrm{SU}(1,1)$ coherent states with which an analog of equation \eqref{scale} can be obtained for both types of spinors. The problem is non-trivial due to the non-compactness of $\mathrm{SU}(1,1)$: since its unitary representations are infinite-dimensional, an identity like \eqref{su2tensor} cannot exist. Coherent states of the discrete series with the intended scaling property had already been found by Perelomov \cite{Perelomov:1986tf}. As far as we are aware, no such states were known for the continous series. 

\section{Coherent states for time-like edges}
\label{sec:weyl}

The ensuing discussion leans heavily on the representation theory of $\mathrm{SU}(1,1)$. The reader is referred to \cite{Bargmann:1946me,ruhl1970lorentz} for the original works, and to \cite{Simao:2024zmv} as the main reference used throughout. In order to simplify notation, we denote the $\mathrm{SU}(1,1)$-invariant pairing as
\begin{equation}
  [u|v]:=\overline{u}^T \sigma_3 v\,, \quad u,v \in \mathbb{C}^2\,,
\end{equation}
and conventionalize that $|v]=\ket{v}$, $[v|=\bra{v}\sigma_3$.

Following the arguments surrounding equation \eqref{scale}, we would like to find reference states $|\chi]\in \mathbb{C}^2$ in the defining representation of $\mathrm{SU}(1,1)$ from which a general normalized Weyl spinor $|z]$ can be obtained. That is, we seek
\begin{equation}
|\chi] \quad \mathrm{s.t.} \quad  g |\chi]\overset{!}{=} \begin{pmatrix}
    z_1 \\ z_2
  \end{pmatrix}\,, \quad g \in \mathrm{SU}(1,1)\,, \quad [z|z]^2=1\,,
\end{equation}
which is clearly possible if $|\chi]=|\pm]$. The $|\pm]$ are eigenstates of $L^3$ in the defining representation. Indeed, the $L^3$ normalized eigenbasis $L^3 f_{k,m}=m f_{k,m}$ of the discrete series\footnote{We restrict for simplicity to the \textit{positive} discrete series, for which $k \in -\frac{\mathbb{N}}{2}$ and $m\in - k  + \mathbb{N}^0$.} $\mathcal{D}^{+}_k$ with spin $k$ can be written as powers of $[\pm|w]$ monomials, with $|w]$ a general Weyl spinor. It reads \cite{Simao:2024zmv}
\begin{equation*}
\begin{gathered}
  f_{k,m}(w_1,w_2)=\frac{(-1)^{k+m}}{\sqrt{\gamma_{k,m}}} [+|w]^{k-m} [-|w]^{k+m}\,, \quad |w]:=\begin{pmatrix}w_1 \\w_2\end{pmatrix} \in \mathbb{C}^2\,,\\
  \gamma_{k,m}:=\frac{\Gamma(-2k-1)\Gamma(1+k+m)}{\Gamma(m-k) }\,,
\end{gathered}
\end{equation*}
the representation acting on such functions as $D^k(g) f_k(w_1, w_2) = f_k(g^ {-1}\begin{psmallmatrix} w_1 \\ w_2 \end{psmallmatrix})$. This suggests that one should consider coherent states constructed from $L^3$ eigenstates, i.e. states of the form $D^{k}(g)f_{k,m_r}$ for some reference $m_r$. In this case
\begin{equation}
  \label{b1}
  \left[D^{k}(g) f_{k,m_r}\right](w_1,w_2)=\frac{(-1)^{k+m_r}}{\sqrt{\gamma_{k,m_r}}} [g\cdot -|w]^{k+m_r} [g \cdot +|w]^{k-m_r}\,,
\end{equation}
and the geometric spinors $|z_{g\pm}]:=|g\cdot \pm]$ (the ones which were put in correspondence with the two-sheeted hyperboloid \eqref{2sheet}) directly figure in the function. 

Looking at the coherent states of equation \eqref{b1}, there exists a distinguished and particularly simple lowest-weight state with $m_r=-k$. Consider then $D^k(g)f_{k,-k}$ which, parametrizing $g=\begin{psmallmatrix} \alpha & \beta \\ \overline{\beta} & \overline{\alpha}\end{psmallmatrix}$, can be expressed using the binomial series as
\begin{align}
\label{minw}
   \left[D^{k}(g) f_{k,-k}\right](1,w)&=\gamma_{k,-k}^{-1/2} (\overline{\alpha} - \beta w)^{2k} \nonumber\\
   &=\gamma_{k,-k}^{-1/2} \sum_{l\geq 0} \binom{2k}{l}\, \overline{\alpha}^{2k-l}(-\beta)^l w^l\,,  
\end{align}
whenever\footnote{It is actually sufficient to consider the action of the maximal compact subgroup generated by $L^3$, in which case the convergence of the binomial series is assured inside the disk $|w|<1$.} $|\beta w/\overline{\alpha}|<1$.  Using the explicit form of the inner product in $\mathcal{D}^+_k$ \cite{Simao:2024zmv}, 
\begin{equation}
  \braket{f,g}_{\mathcal{D}^{+}_k}:=   \int_{D^1} {\underbrace{i(2\pi)^{-1}\diff w \wedge \diff \overline{w} \, (1-|w|^2) }_{:= \diff  \omega_w}}^{-2k-2} \overline{f}(1,w) g(1,w)\,,
\end{equation}
one has for the pairing of two such coherent states that
\begin{align}
  \langle D^{k}(g')&f_{k,-k},D^k(g)f_{k,-k} \rangle \nonumber\\
  &=\frac{\alpha'^{2k}\overline{\alpha}^{2k}}{\gamma_{k,-k}\pi}\sum_{l,t\geq 0}\binom{2k}{t}\binom{2k}{l}\left(\frac{-\overline{\beta}'}{\alpha'}\right)^t\left(\frac{-\beta}{\overline{\alpha}}\right)^l  \int_{D^1} \diff  \omega_w \, \overline{w}^t w^l \nonumber \\
  &= \sum_{l\geq 0} \binom{2k}{l}^2 \binom{-2k+l-1}{l}^{-1} (\alpha'\overline{\alpha})^{2k-l} (\overline{\beta}' \beta)^l \nonumber \\
  &= (\alpha' \overline{\alpha}- \overline{\beta}'\beta)^{2k}\,.
\end{align}
Hence the following identity involving the defining representation holds,
\begin{align}
\label{2pointsl}
  \braket{k, -k|D^{k\dagger}(g')D^k(g)|k,-k}&=\braket{-|g'^{\dagger} \sigma_3 g|-}^{2k} \nonumber \\
  &=[z_{g'}| z_g]^{2k}\,,
\end{align}
having set $|z_{g}]:=g|-]$. Our choice of coherent states thus satisfies the requirements outlined earlier: their inner product amounts to a pairing of $H^+$ Weyl spinors, scaling with the power of spin $k$. These are the original states of \cite{Perelomov:1986tf}, also employed in the CH spin-foam amplitude \cite{Conrady:2010kc}.

\section{Coherent states for space-like edges}
\label{sec:maj}
We now turn our attention to the continuous series. Analogously to the above, we require that an $\mathrm{SU}(1,1)$ action on a pair of reference states $|\chi_i]$ yields a Majorana pair,
\begin{equation}
 |\chi_i] \quad \mathrm{s.t.} \quad  g |\chi_i]\overset{!}{=} \begin{pmatrix}
    z_i \\ \overline{z_i}
  \end{pmatrix}\,, \quad g \in \mathrm{SU}(1,1)\,, \quad [z^2|z^1]^2=-1\,.
\end{equation}
It is straightforward to verify that $|\chi_1]=|l^+]$ and $|\chi_2]=i|l^-]$ solve the problem, having defined
\begin{equation}
  |l^+]:=\frac{1}{\sqrt{2}} \left(|+] + |-]\right)\,, \quad   |l^-]:=\frac{1}{\sqrt{2}} \left(|+] - |-]\right)\,.
\end{equation}
The vectors $|l^\pm]$ are not $L^3$ eigenstates, but rather eigenstates of $K^2:=-i\sigma_1/2$ in the defining representation. It is thus reasonable to expect that an identity similar to that of equation \eqref{2pointsl} may hold, but seemingly requiring coherent states constructed from $K^2$. It is known \cite{Lindblad:1969zz} that $K^2$, being a non-compact operator with continuous spectrum, does not have eigenstates in the representation spaces\footnote{The notation $\mathcal{C}^\delta_j$ stands for the continuous series Hilbert space of spin $j=-1/2+is$, $s>0$. The eigenstates of $L^3$ are labeled by $m\in \delta + \mathbb{Z}\,, \;\delta\in\{0,\frac{1}{2}\}$; see e.g. \cite{Bargmann:1946me,ruhl1970lorentz,Simao:2024zmv}.} $\mathcal{C}^\delta_j$; moving forward with the construction requires first pointing out a few facts about generalized eigenstates. 

\subsection{Gelfand triple in $\mathcal{C}^\delta_j$}
\label{sec:gelfand}
We follow the treatment of the subject given by Lindblad in \cite{Lindblad:1969zz}, remarking that our conventions match exactly those provided in that work. 

Given the Hilbert space $\mathcal{C}^\delta_j$, it is possible to define a dense subspace $\mathcal{D}$ of ``rapidly decreasing sequences'' 
\begin{equation}
  \mathcal{D}=\left\{ \sum_m c_m \ket{j,m}\; \Bigl|\; \forall n\in \mathbb{N}\,, \; \lim_{|m| \to \infty} m^n c_m = 0  \right\}\,,
\end{equation}
with a certain topology cf. \cite{Lindblad:1969zz}. There is then a Gelfand triple $\mathcal{D}\subset \mathcal{C}^\delta_j \subset \mathcal{D}'$, where $\mathcal{D}'$ is the space of continuous functionals on $\mathcal{D}$ and $\mathcal{C}^\delta_j$ (identified with its dual) is dense in $\mathcal{D}'$.
The functional-analytical nuances of the construction are such that a spectral theorem can be applied to self-adjoint operators with continuous spectra on such a triple. For the purposes of this work it is sufficient to state that the nuclear spectral theorem guarantees that $K^2$ (by virtue of being self-adjoint and continuous in $\mathcal{D}$ and leaving it invariant) has a complete set of generalized eigenvectors in $\mathcal{D}'$. This is meant in the sense that
\begin{equation}
  F_{\lambda,\sigma}({K^2}^\dagger \psi_m)= \lambda \braket{j, m| j, \lambda}\,, \quad F_{\lambda,\sigma} \in \mathcal{D}'\,, \quad \psi_m \in \mathcal{D},
\end{equation}
\begin{equation}
\label{k2complete}
  \braket{\psi,\varphi}_{\mathcal{C}^\delta_j}= \sum_\sigma \int \diff \lambda \, \braket{\psi|j, \lambda, \sigma} \braket{j, \overline{\lambda}, \sigma|\varphi}\,,
\end{equation}
with $\sigma$ standing for the degeneracy of the distribution at $\lambda$. It is crucial for the following to note that $\lambda \in \mathbb{C}$ is not required to be real since - as pointed out by Lindblad himself - while $K^2$ is self-adjoint in $\mathcal{D}$ its extension to $\mathcal{D}'$ is not. That a complex-conjugated eigenvalue $\overline{\lambda}$ is necessary in the completeness relation \eqref{k2complete} follows however from the self-adjoint property of $K^2$ in $\mathcal{C}^\delta_j$, 
\begin{align}
\braket{K^{2\dagger} \psi, \varphi}_{\mathcal{C}^\delta_j}&=\sum_\sigma \int \diff \lambda \, F_{\lambda,\sigma}(K^{2\dagger} \psi) \overline{F_{\overline{\lambda},\sigma}(\varphi)}\nonumber \\
&= \sum_\sigma \int \diff \lambda\, F_{\lambda,\sigma}(\psi)\, \lambda \, \overline{F_{\overline{\lambda},\sigma}(\varphi)} \nonumber\\
&= \braket{\psi, K^{2\dagger} \varphi}_{\mathcal{C}^\delta_j}\,.
\end{align}
Since we will be interested in determining matrix coefficients of the type \linebreak $\braket{j, \overline{\lambda'}, \sigma'|D^j(g)|j, \lambda, \sigma}$, it is important to note that such objects may not be by themselves well-defined; they are generally to be understood as distributions, i.e.
\begin{equation}
  F_{\lambda,\sigma}(D^{j\dagger}(g)\psi)=\sum_{\sigma '} \int \diff \lambda' \, \braket{\psi|j,\lambda',\sigma'}\braket{j, \overline{\lambda'}, \sigma'|D^j(g)|j, \lambda, \sigma}\,. 
\end{equation}

\subsection{The generalized $K^2$ eigenbasis}

With the above preliminaries established, we proceed to finding a complete set of generalized eigenstates of $K^2$. Inspired by the discussion of section \ref{sec:weyl}, we consider powers of $[l^\pm|w]$ monomials, with $|w]$ a general Majorana spinor, and define\footnote{$z^w:=e^{w \ln z}$ for $z,w \in \mathbb{C}$ is defined in terms of the principal branch $\mathrm{arg} z \in [-\pi,\pi)$ of the logarithm.}
\begin{equation*}
  \label{K2eigen}
  \begin{gathered}
  f_{j,\lambda}^\sigma(w)= \alpha_{j}\left|[l^-|w]\right|^{\overline{j}-i\lambda} \,  \left|[l^+|w]\right|^{\overline{j}+i\lambda} \mathrm{sgn}^\sigma \Im \left([l^+|w][l^-|w]\right)  \mathrm{sgn}^{2\delta} \Re\left( [l^-|w] \right) \,, \\
  |w]:= \begin{pmatrix} w \\ \overline{w} \end{pmatrix} \in \mathbb{C}^2\,, \quad  \alpha_{j}:=2^j\,.
  \end{gathered}
\end{equation*}
These functions are homogeneous of degree $-2j-2$, and their restriction to the circle $|z|=1$ yields (upon setting $\theta=\mathrm{arg}\, z$)
\begin{equation}
\label{K2basis}
  f_{j,\lambda}^\sigma(\theta)= \frac{1}{2}|\cos\theta|^{\overline{j}-i\lambda}\, |\sin\theta|^{\overline{j}+i\lambda} \,\mathrm{sgn}^\sigma\left[ \cos\theta\sin\theta\right]\, \mathrm{sgn}^{2\delta} \left[\cos\theta \right]\,.
\end{equation}
Appealing to the spinor representation of the Casimir $Q$ and $K^2$ operators \cite{Simao:2024zmv},
\begin{equation}
  \begin{gathered}
    Q=\frac{1}{4} \left[ 2(w \partial+\overline{w} \overline{\partial}) + (w \partial+\overline{w} \overline{\partial}) ^2 \right] \,, \\
    K^2=\frac{i}{2}\left(w  \overline{\partial}+\overline{w} \partial \right)\,, 
  \end{gathered}
\end{equation}
it is straightforward to show
\begin{equation}
\begin{gathered}
Q  f^\sigma_{j,\lambda}(w) = j(j+1) f^\sigma_{j,\lambda}(w) \,, \\
  K^2 f^\sigma_{j,\lambda}(w) = \lambda f^\sigma_{j,\lambda}(w) \,, 
  \end{gathered}
\end{equation}
confirming that these are indeed eigenstates. The $\mathrm{sgn}$ factors in $f^\sigma_{j,\lambda}(w)$ were introduced in order to control its parity behavior as the argument moves around the circle $\theta \in [0, 2\pi)$. It holds that
\begin{equation}
\begin{gathered}
  P  f_{j,\lambda}^\sigma(w):= f_{j,\lambda}^\sigma(\overline{w}) = (-1)^\sigma  f_{j,\lambda}^\sigma(w)\,, \quad \sigma\in\{0,1\}\,,\\
  f_{j,\lambda}^\sigma(\theta+\pi)=(-1)^{2\delta}  f_{j,\lambda}^\sigma(\theta)\,, \quad \delta\in \{0,\frac{1}{2}\}\,,
  \end{gathered}
\end{equation}
so that the states diagonalize $P^2=\mathds{1}$ and satisfy the periodicity property of $\mathcal{C}^\delta_j$, just as required by Lindblad in \cite{Lindblad:1969zz}. 

From the definition of the inner product in $\mathcal{C}_j^\delta$ \cite{Simao:2024zmv}
\begin{equation}
  \label{preinnersl2}
    \braket{f,g}_{\mathcal{C}_j^\delta}=\frac{1}{2\pi}\int_{S^1} \diff \theta \; \overline{f}(\theta) g(\theta)\,,
  \end{equation}
and the $L^3$ orthonormal eigenbasis obtained in \cite{Simao:2024zmv}
\begin{equation}
  f_{j,m}(w)= \left(\frac{\Gamma(m-\overline{j})}{\Gamma(m-j)}\right)^{\frac{1}{2}}  w^{-j-1-m} \overline{w}^{-j-1+m} \,,
\end{equation}
one can explicitly derive the integral identity
\begin{align}
  \braket{j,m|j, \lambda, \sigma} = \frac{1}{2\pi} \left(\frac{\Gamma(m-j)}{\Gamma(m-\overline{j})} \right)^{\frac{1}{2}} \int_0^{\pi/2}&\left(e^{2i\theta m}+(-1)^\sigma e^{-2i\theta m}\right) \nonumber \\
  \label{intid}
   &\quad  \cdot (\cos{\theta})^{\overline{j}-i\lambda} (\sin{\theta})^{\overline{j}+i\lambda}\, \diff \theta \,,
\end{align}
recovering a result of \cite{Lindblad:1969zz}. Making use of equation \eqref{intid} it is possible to verify the completeness and orthogonality of $f^\sigma_{j,\lambda}(w)$ states, 
\begin{equation}
  \begin{gathered}
  \sum_\sigma \int_{\mathbb{R}+ix} \diff\lambda \,  \braket{j,m|j, \lambda, \sigma} \braket{j, \overline{\lambda}, \sigma|j,n}=\delta_{mn}\,, \quad x \in \mathbb{R}\,, \\
  \sum_m  \braket{j, \overline{\lambda'}, \sigma'|j,m} \braket{j,m|j, \lambda, \sigma} = \delta(\lambda-\lambda')\,, \quad \Im \lambda =\Im \lambda'\,.
  \end{gathered}
\end{equation}
Crucially, completeness holds whenever $\lambda$ is integrated over a real line displaced by a fixed imaginary factor $ix$. In short, there is a family of orthonormal bases
\begin{equation}
  \left\{\, \ket{j, \lambda+ix, \sigma} \, | \, \lambda \in \mathbb{R}\,, \sigma \in \{0,1\}\right\}_{x \in \mathbb{R}}
\end{equation}
indexed by a real number $x$. Observe however that the $\sigma$ labels are not orthogonal among themselves.

\subsection{A proposal for $K^2$ coherent states}
\label{sec:k2reg}
In the case of the discrete series of section \ref{sec:weyl}, there was a suggestive choice of reference state with which to define coherent states: the lowest-weight state, for which the corresponding function \eqref{minw} is particularly simple. The fact that the continuous series does not terminate in either direction of $\lambda$ invalidates applying the same criterion. An alternative requirement, which we have already suggested in \cite{Simao:2021qno}, is to consider those states which minimize the variance of the generators $F^i:=(L^3, K^1, K^2)$, appropriately defined to account for the fact that $\lambda$ may be complex. We thus propose to consider those states which minimize
\begin{equation}
  \braket{\Delta |F^i|}:=\braket{F^i F_i}-\braket{F^i}\overline{\braket{F_i}}=-s^2-\frac{1}{4}+|\lambda|^2\,,
\end{equation}
i.e. those states lying in the circle $|\lambda|^2=s^2+\frac{1}{4}$. Our choice is to restrict to the complex value\footnote{Note the formal similarity to the discrete series case $m=-k$ of section \ref{sec:weyl}, compared to the $\lambda=s$ prescription of \cite{Conrady:2010kc}.} $\lambda= i j$ (recall $j=-1/2+is$ labels the continuous series), and it will be shown that pairings of the type $\braket{j,\overline{ij},\sigma| D^j(g)|j,ij,\sigma}$ satisfy the desired scaling properties. Since the presence of a conjugated eigenvalue introduces an asymmetry to the pairing, the object $\braket{j,ij,\sigma| D^j(g)|j,\overline{ij},\sigma}$ will also be considered.

In determining matrix coefficients it turns out to be enough to consider the maximal subgroup generated by $L^3$. The general expression for \newline $\braket{j,\overline{\lambda},\sigma| D^j(e^{-i \alpha L^3})|j,\lambda',\sigma'}$ has already been derived in \cite{Lindblad:1969zz} for arbitrary complex eigenvalues. The result is a well-defined meromorphic function of $\lambda$ and $\lambda'$, except when $\Delta\lambda:=\lambda-\lambda'=0$. Settling on a particular choice of $\lambda$ for all coherent states requires addressing this singularity. Fortunately, the matrix coefficients can easily be regularized\footnote{The proposed regularization cannot be implemented by a $\ket{j,\lambda+\epsilon,\sigma}$ displacement of the reference state. We thus interpret our regularization as a modification of the function of matrix coefficients itself, and not of the boundary states.} by taking  $\Gamma(\pm i\Delta\lambda) \mapsto \Gamma(\pm [i\Delta\lambda +\epsilon])$ in the original formula \cite{Lindblad:1969zz}. We therefore define
\begin{align}
  \label{coef}
& \braket{j,\overline{\lambda},\sigma| D^j(e^{-i \alpha L^3})|j,\lambda',\sigma'}_{\mathrm{reg}}= \nonumber\\
& \lim_{\epsilon \to 0} \frac{1}{2\pi}  \Biggl\{ \frac{\Gamma(\frac{-\overline{j}+\sigma-i\lambda}{2}) \Gamma(\frac{-j+\sigma'+i\lambda'}{2})}{\Gamma(\frac{-j+\sigma+i\lambda}{2}) \Gamma(\frac{-\overline{j}+\sigma'-i\lambda'}{2})} \Gamma(i\Delta \lambda + \epsilon)\psi_-(\alpha) \nonumber  \\
 & \hspace{0.5cm} +(-1)^{2\delta}  \frac{\Gamma(\frac{-\overline{j}+2\delta+(-1)^{2\delta}\sigma+i\lambda}{2}) \Gamma(\frac{-j+2\delta+(-1)^{2\delta}\sigma'-i\lambda'}{2})}{\Gamma(\frac{-j+2\delta+(-1)^{2\delta}\sigma-i\lambda}{2}) \Gamma(\frac{-\overline{j}+2\delta+(-1)^{2\delta}\sigma'+i\lambda'}{2})} \Gamma(-i\Delta \lambda - \epsilon)\psi_+(\alpha) \Biggr\} \nonumber \\
 & \hspace{7.5cm} \cdot \, \cos \frac{\pi}{2}(i\Delta \lambda+\sigma-\sigma')\,,
\end{align}
where $\alpha\in [-\pi,\pi]$ and $\psi_\pm(\alpha)$ takes the form
\begin{align}
  \psi_\pm(\alpha)=\cos& \left(\frac{\alpha}{2}\right)^{-2j-2}\left|2\tan \frac{\alpha}{2} \right|^{\pm i\Delta\lambda} \nonumber \\
  & \cdot\,  {}_2F_1\left(j+1\pm i\lambda, j+1\mp i\lambda',1\pm i\Delta \lambda ;-\tan^2 \frac{\alpha}{2}\right)\mathrm{sgn}^{\sigma-\sigma'}\alpha\,.
\end{align}
Resorting to a well-known identity for the hypergeometric function ${}_2F_1$ with repeated coefficients, and setting $\lambda= ij$, it is straightforward to see that 
\begin{align}
\label{r1}
  \bra{j,\overline{ ij},\sigma}  D^j(e^{-i \alpha L^3})& \ket{j,ij,\sigma}^{\delta=0}_{\mathrm{reg}} \nonumber\\
  &= \lim_{\epsilon \to 0} \frac{\Gamma(\epsilon)+ \Gamma(-\epsilon) }{2\pi} \cos\left(\frac{\alpha}{2}\right)^{-2j-2} \left(1+\tan^2 \frac{\alpha}{2}\right)^{2j+1} \nonumber\\
  &= -\frac{\gamma}{\pi} \cos^{2j} \frac{\alpha}{2} \,, 
\end{align}
with $\gamma$ the Euler-Mascheroni constant (not to be confused with the Immirzi parameter). Likewise\footnote{Observe that $\overline{j}=-2j-2$.}
\begin{equation}
  \label{reg}
  \bra{j,ij,\sigma}  D^j(e^{-i \alpha L^3}) \ket{j,\overline{ij},\sigma}^{\delta=0}_{\mathrm{reg}}  = -\frac{\gamma}{\pi} \cos^{-2j-2} \frac{\alpha}{2}\,,
\end{equation}
which can be shown either from complex-conjugation of \eqref{reg} or by direct computation from equation \eqref{coef}. 

One must take care in generalizing the matrix coefficients to an arbitrary $g\in \mathrm{SU}(1,1)$, since intermediate computation steps may diverge; one ought to appeal to equation \eqref{r1} as much as possible. To that end note first that 
\begin{equation}
\label{semi1}
  \left[\sum_m  e^{-i \alpha m}\braket{j,\overline{ij},\sigma|j,m}\braket{j,m|j,ij,\sigma}\right]_{\mathrm{reg}}  = -\frac{\gamma}{\pi} \cos^{2j} \frac{\alpha}{2} \,,
\end{equation}
by virtue of completeness of $L^3$ eigenstates $\ket{j,m}$ in $\mathcal{C}^0_j$. Moreover, making use of the  parametrization\footnote{We follow the convention that $L^3=\sigma_3/2$, $K^1=i\sigma_2/2$ and $K^2=-i\sigma_1/2$, as in \cite{Simao:2024zmv}.} $g=e^{-i\alpha L^3}e^{-i tK^1}e^{-i u K^2}$ (see e.g. \cite{Conrady:2010sx}), it must be that 
\begin{align}
\label{param}
  \bra{j,\overline{ij},\sigma} &D^j(g)\ket{j,ij,\sigma}_{\mathrm{reg}} \nonumber \\
  &=e^{ju} \left[\sum_m e^{-i \alpha m} \braket{j,\overline{ij},\sigma|j,m} \braket{j,m|D^j(e^{-i t K^1})|j,ij,\sigma}\right]_{\mathrm{reg}}\,.
\end{align}
The matrix elements of the subgroup generated by $K^1$ were also determined by Lindblad in a subsequent paper \cite{Lindblad:1970tv}, from where the identity
\begin{equation}
\label{semi2}
  \braket{j,m|D^j(e^{-i t K^1})|j,ij,\sigma}=\braket{j,m|j,ij,\sigma}\left(1+i\sinh t\right)^\frac{j+m}{2} \left(1-i\sinh t\right)^\frac{j-m}{2}
\end{equation}
can be obtained. Thus, plugging-in equations \eqref{semi1}, \eqref{semi2} in \eqref{param}, and setting $\phi:=\arg\left(\cosh \frac{t}{2}+ i\sinh\frac{t}{2} \right)$,
\begin{align}
  &\bra{j,\overline{ij},\sigma} D^j(g)\ket{j,ij,\sigma}_{\mathrm{reg}} \nonumber \\
  &= \left| \cosh \frac{t}{2}+i\sinh\frac{t}{2}\right|^{2j} e^{ju}  \left[ \sum_m e^{-i m (\alpha - 2\phi)} \braket{j,\overline{ij},\sigma|j,m}\braket{j,m|j,ij,\sigma}\right]_{\mathrm{reg}} \nonumber \\
  &=-\frac{\gamma}{\pi} \left| \cosh \frac{t}{2}+i\sinh\frac{t}{2}\right|^{2j} \cos^{2j} \left(\frac{\alpha}{2}-\phi\right)e^{ju} \nonumber \\
  &=  -\frac{\gamma}{\pi}\braket{l^-|g|l^-}^{2j}\,.
\end{align} 
One moreover finds
\begin{equation}
  \bra{j,ij,\sigma} D^j(g)\ket{j,\overline{ij},\sigma}_{\mathrm{reg}} =  -\frac{\gamma}{\pi}\braket{l^+|g|l^+}^{-2j-2} \,,
\end{equation}
which shows that a different ordering of the matrix coefficients amounts to a conjugation of the spin, together with the substitution $\ket{l^-}\mapsto \ket{l^+}$. 

Finally, defining $|z^+_{g}]:=g|l^+]$ and $|z^-_g]:= ig|l^-]$, we arrive at the principal result of this section:
\begin{align}
\label{r2}
  \braket{j,\overline{ij},\sigma|D^{j\dagger}(g') D^j(g)|j,ij,\sigma}_{\mathrm{reg}}&=-\frac{\gamma}{\pi} \braket{l^+|{g'}^\dagger\sigma_3 g|l^-}^{2j}\nonumber \\
  &= -\frac{\gamma}{\pi} \left(-i [z^+_{g'}|z^-_g]\right)^{2j}\,,
\end{align} 
together with 
\begin{align}
  \label{r3}
    \braket{j,ij,\sigma|D^{j\dagger}(g') D^j(g)|j,\overline{ij},\sigma}_{\mathrm{reg}}&=-\frac{\gamma}{\pi}\braket{l^-|{g'}^\dagger\sigma_3 g|l^+}^{-2j-2}\nonumber \\
    &= -\frac{\gamma}{\pi} \left(i [z^-_{g'}|z^+_g]\right)^{-2j-2}\,.
  \end{align} 
As intended, the pairing of coherent states scales with the spin label, and it relates to Majorana spinors which can be put in correspondence with $H^{\mathrm{sl}}$ as per equation \eqref{1sheet}.   

A comment on the regularization procedure leading to equations \eqref{r2} and \eqref{r3} is due. It is undeniable that regularizing a diverging object is less of a science than it is an art: the procedure is not unique, and strictly speaking one can only say with certainty that without it the matrix coefficients are undefined. Still, the circumstance that this specific regularization of generalized eigenstates is particularly simple, and that it leads - as per \eqref{r2} - to the same qualitative behavior as that of the discrete series coefficients \eqref{2pointsl} (and indeed the $\mathrm{SU}(2)$ ones) lends credence to the choice made. There is a parallel to be made with the well-known Feynman $i\epsilon$ regularization of the generalized eigenfunctions of the momentum operator in QFT, ultimately legitimized by the empirical success of the theory. For what concerns spin-foams an appeal to plausibility will have to suffice.

\section{A 3d Lorentzian vertex amplitude}

Having found appropriate coherent states for both space- and time-like boundaries, we proceed in explicitly defining the model. Three basic ingredients are needed. First observe that there exists a Plancherel formula for smooth compactly-supported functions on $\mathrm{SU}(1,1)$ \cite{Pukanszky:1963,Takahashi:1961,ruhl1970lorentz}, leading to a harmonic expansion of the Dirac delta distribution as
\begin{align}
\label{delta}
  \delta(g)=\sum_{\delta=0,\frac{1}{2}} \int_{-\infty}^\infty \diff s\, s\tanh^{1-4\delta} (\pi s) \, &\mathrm{Tr}\left[D^{j(\delta)}(g)\right] \nonumber \\
  +\sum_{q=\pm} \sum_{2k=-1}^{-\infty}& (-2k-1)\mathrm{Tr}\left[D^{k(q)}(g)\right]\,,
\end{align}
where the notation\footnote{In equation \eqref{delta}, the letter $q$ labels the positive $q=+$ or negative $q=-$ discrete series of $\mathrm{SU}(1,1)$ unitary representations \cite{Simao:2024zmv}.} and conventions used throughout the text have been kept. 
\vspace*{-1\baselineskip}
\begin{table}
\centering
\caption{\small Cells of a 2-complex dual to a 3d triangulation} 
\vspace{1ex}
\small \begin{tabular}{l l} 
\toprule 
\multicolumn{1}{c}{2-complex $\Delta^*$} & \multicolumn{1}{c}{ triangulation $\Delta$ }\\ \midrule
vertex $v$ & tetrahedron $\tau$ \\  
edge $e$ & triangle $t$ \\ 
face $f$ & triangle-edge $\epsilon$  \\ 
\bottomrule 
\end{tabular}
\label{tab:dual3}
\end{table}

Secondly recall that - as a purely topological theory - the tetradic action for $2+1$ gravity agrees with the 3d version of unconstrained $BF$ theory,
\begin{align}
S[\theta, A]=\int_M (\star\theta)_{IJ} F^{IJ} \; \leftrightarrow\; S[B,A]=\int_M B_{IJ} F^{IJ}\,,
\end{align}
where $I=0,1,2$, $M$ is a smooth 3-manifold without boundary, and the gauge group is $\mathrm{SU}(1,1)$. The usual spin-foam quantization procedure (see e.g. \cite{Baez:1999sr}) can then be followed: introducing a triangulation $\Delta$ of $M$, together with its dual complex $\Delta^*$ (as per table \ref{tab:dual3}), one obtains the \textit{formal} partition function
\begin{align}
\label{partition}
    Z&(\Delta^*)= \sum_{\delta \rightarrow f}\int_{j\rightarrow f}\diff s\, \prod_f \, \left[s_f \tanh^{1-4\delta_f}\left(\pi s_f\right)\right] \nonumber \\
    & \hspace{3cm}  \cdot \, \mathrm{Tr}_f\,\Bigl[\prod_e \Bigl(\int \diff g_e   \prod_{f \,\mathrm{s.t.}\, e\in \partial f} D^{j_f(\delta_f)}(g_e) \Bigr)\Bigr] \nonumber\\
    &+ \sum_{q \rightarrow f} \sum_{k \rightarrow f} \prod_f \, \left(-2k_f-1\right)\, \mathrm{Tr}_f\,\left[\prod_e \left(\int \diff g_e  \prod_{f \,\mathrm{s.t.}\, e\in \partial f} D^{k_f(q_f)}(g_e) \right)\right]\,.
\end{align} 
The notation $\mathrm{Tr}_f$ means that the matrix coefficients are to be contracted along the boundary of the face $f \in \Delta^*$ (clarifying diagrams can be found in \cite{Perez:2012wv}). Any concerns regarding convergence or computational viability are for now to be boldly ignored. 

Thirdly, it is necessary to pick a basis with which to take traces. Although the trace is invariant under basis transformations, the chosen basis will induce specific boundary states once a vertex amplitude with boundary is induced from the partition function. Since the coherent states proposed in sections \ref{sec:weyl} and \ref{sec:maj} can be shown to satisfy completeness
\begin{equation}
\label{cohcompdis}
 \int\diff g\; D^{k(q)}(g)\ket{k,-q k}\bra{k,-q k} D^{k(q)\dagger}(g)=\frac{\mathds{1}_{k(q)}}{-2k-1}\,,
\end{equation}
\begin{equation}
\label{cohcompcont}
  \int\diff g\; D^{j(\delta)}(g)\ket{j,\lambda, \sigma}\bra{j',\overline{\lambda}',\sigma} D^{j(\delta)\dagger}(g)=\frac{\mathds{1}_{j(\delta)}\delta(j-j')\delta(\lambda-\lambda')}{ s \tanh^{1-4\delta}\pi s}\,,
\end{equation}
they may be used as a basis for the trace operation\footnote{The two Dirac deltas in equation \eqref{cohcompcont} reflect the facts that 1) the matrix coefficients of the continuous series are not square-integrable \cite{Bargmann:1946me}, and 2) the states $\ket{j,\lambda,\sigma}$ are only \textit{generalized} eigenstates.}. The relevant identities $\mathds{1}_{k(q)}$ and $\mathds{1}_{j(\delta)}$ can be inserted in the partition function \eqref{partition}, from where a vertex amplitude with boundary coherent states can be extracted. 

Some more notation is needed in order to express the amplitude. Set $\ket{qk,g}:=D^{k(q)}(g)\ket{k, -qk}$, $\ket{j,g}:=D^{j(0)}(g)\ket{j, ij, 0}$ and $|j,g]:=D^{j(0)}(g)\ket{j, \overline{ij}, 0}$ for the remainder of this text. Introduce the diagram
\begin{equation}
\label{d1}
\!{}^{ n}\scalebox{0.65}{\input{Pics/braket.tikz}}^{ n'}:= d_k \braket{qk,n|D^{k(q)}(g)|qk,n'}\,, \quad d_k=-2k-1\,,
\end{equation}
for a pairing of time-like coherent states. The asymmetry of space-like coherent states allows for two inequivalent pairings (denoted $\oplus$ and $\ominus$),
\begin{equation}
\label{d2}
  \!{}^{ n}\underset{\oplus}{\scalebox{0.65}{\input{Pics/braket_black1.tikz}}}^{ n'}:=d_s \,  \mathcal{C}_{n,gn'} \; [j,n|D^{j(0)}(g)|j,n'\rangle \,, \quad d_s= s \tanh \pi s \,,
  \end{equation}
 \begin{equation}
 \label{d3}
   \!{}^{ n}\underset{\ominus}{\scalebox{0.65}{\input{Pics/braket_black2.tikz}}}^{ n'}:=d_s \, \mathcal{C}_{n,gn'} \;  \langle j,n|D^{j(0)}(g)|j,n'] \,.
\end{equation}
The term $\mathcal{C}_{n,n'}$ is a function of the boundary data
\begin{equation}
\label{constr}
   \mathcal{C}_{n,n'}:= e^{s [l^+_n|l^+_{n'}]^2}\,, \quad \ket{l^\pm_n}:=n\ket{l^\pm}\,,
\end{equation}
which we have appended to the space-like pairings ex post facto. Its inclusion will prove fundamental in guaranteeing a well-behaved semi-classical limit, as it corresponds to an otherwise absent Gaussian implementation of a gluing constraint between the edges $n$ and $n'$. We will come back to this point in the context of asymptotic analysis.

The model we wish to propose now follows. Pick $12$ group elements $n_{ab}\in\mathrm{SU}(1,1)$ and $6$ spin labels $j_{ab}=j_{ba}$ or $k_{ab}=k_{ba}$ (and in that case also $q_{ab}=q_{ba}$) as boundary data, one for each edge of a tetrahedron. The vertex amplitude is constructed from a convolution of the diagrams \eqref{d1} to \eqref{d3}, following the combinatorics of a tetrahedron. For example, the amplitude for a tetrahedron with all edges time-like is given by
\begin{align}
   \scalebox{0.35}{ \input{Pics/vertexsu2.tikz}}&=\int \prod_{a=1}^3 \diff g_a\, \prod_{a< b} d_{k_{ab}} \braket{qk_{ab},n_{ab}|D^{k_{ab}(q)_{ab}}(g_a)^\dagger D^{k_{ab}(q)_{ab}}(g_b)|qk_{ab},n_{ba} } \nonumber \\
   &= \int \prod_{a=1}^3 \diff g_a\, \prod_{a< b} d_{k_{ab}} \braket{(-q)_{ab}|g_a^\dagger \sigma_3 g_b|(-q)_{ba}}^{2k_{ab}}\,, \label{A3dtl}
\end{align}
where $\ket{\pm_{ab}}:=n_{ab}\ket{\pm}$. The amplitude for a tetrahedron with all edges space-like, and using only  pairings of type $\oplus$, reads
\begin{align}
  &\scalebox{0.35}{ \input{Pics/vertexsu2_black.tikz}}^\oplus =\int \prod_{a=1}^3 \diff g_a\, \prod_{a< b} d_{s_{ab}}\, \mathcal{C}_{g_a n_{ab},g_b n_{ba}}\; [j_{ab},n_{ab}|D^{j_{ab}(0)}(g_a)^\dagger D^{j_{ab}(0)}(g_b)|j_{ab},n_{ba} \rangle  \nonumber \\
  &= \int \prod_{a=1}^3 \diff g_a\, \prod_{a< b} \left(-\frac{\gamma d_{s_{ab}}}{\pi}e^{s_{ab} \braket{l^+_{ab}|g_a^\dagger \sigma_3 g_b |l^+_{ba}}^2}\right) \braket{l^+_{ab}|g_a^\dagger \sigma_3 g_b|l^-_{ba}}^{-1+2is_{ab}}\,. \label{A3dsl}
\end{align}
A general vertex with space- and time-like edges involves combinations of all types of pairings. On all amplitudes we set $g_4=\mathds{1}$ in order to regularize the Haar integral, as usual. 

\subsubsection*{A note on the space-like dichotomy}

Notice that different combinations of $\oplus$ and $\ominus$ space-like pairings result in bona-fide different vertex amplitudes, so that a selection criterion ought to be introduced. It is quite tempting to interpret the binary classification of space-like edges as a causal ordering on space-like \textit{wedges}, just as has earlier been proposed in \cite{Livine:2002rh, Bianchi:2021ric}; in that context, the binary choice $\pm$ corresponds to the relative time-orientation of triangle normals, and the amplitude associated to a tetrahedron with a combination of $\pm$ signs is semiclassicaly subleading \cite{Bianchi:2021ric}. Such an interpretation for our model is however flawed, since - as will be argued in the following section - not only is the semiclassical amplitude of a tetrahedron with $\oplus$ and $\ominus$ edges not subleading, the asymptotic formula fails to recover the Regge action for the associated geometry. The amplitude only displays the correct semiclassical behavior when a coherent assignment of space-like edge orientations is made. This constitutes a criterion, and our choice is to settle on $\oplus$-type amplitudes.

\section{Asymptotic formula}
\label{sec:asy}

The introduction of \eqref{A3dsl} as a vertex amplitude for space-like boundaries is partially predicated on  its desirable asymptotic behavior; it remains to investigate it. We shall follow protocol \cite{Barrett:2009mw, Kaminski:2017eew, Liu:2018gfc, Simao:2021qno} and resort to a stationary phase approximation of the amplitude for large spins $s$. For ease of presentation we restrict to a fully $\oplus$-type amplitude, and simply comment on the result for mixed dichotomies.

The vertex amplitude with uniformly scaled spins $\Lambda s_{ab}$ can be rewritten as an exponential integral,
\begin{equation}
  \label{amp2}
   \scalebox{0.35}{ \input{Pics/vertexsu2_black.tikz}}^{\oplus}_\Lambda= \left(\frac{\gamma \Lambda}{\pi}\right)^6 \int \prod_{a=1}^3 \diff g_a\, \prod_{a< b} \frac{s_{ab}\tanh (\pi \Lambda s_{ab})}{ \braket{l^+_{ab}|g_a^\dagger \sigma_3 g_b|l^-_{ba}}}e^{\Lambda S_{ab}}\,,
\end{equation}
with an action given by
\begin{equation}
  S_{ab}= 2is_{ab}\ln \braket{l^+_{ab}|g_a^\dagger \sigma_3 g_b|l^-_{ba}}+ s_{ab}\braket{l^+_{ab}|g_a^\dagger \sigma_3 g_b|l^+_{ba}}^2\,.
\end{equation}
By Hörmander's theorem \cite[Th. 7.7.5]{Hormander2003}, when $\Lambda \to \infty$ the integral is dominated by stationary contributions $\delta_g S_{ab}=0$ with maximal real part. Observe to that end that $\braket{l^+_{ab}|g_a^\dagger \sigma_3 g_b|l^-_{ba}}\in \mathbb{R}$, so that the maximum of $\Re S_{ab}$ is attained at 
\begin{equation}
  \Re S_{ab}=0 \quad \Leftrightarrow \quad \braket{l^+_{ab}|g_a^\dagger \sigma_3 g_b|l^+_{ba}}=0\; \wedge \; \braket{l^+_{ab}|g_a^\dagger \sigma_3 g_b|l^-_{ba}}>0\,, 
\end{equation}
which implies $g_b \ket{l^+_{ba}}=\vartheta_{ab}g_a \ket{l^+_{ab}}$; acting with the $\mathrm{SU}(1,1)$-compatible real structure $R$ \cite{Simao:2024zmv} shows that $\vartheta_{ab} \in \mathbb{R}$, and the second condition sets $\vartheta_{ab}>0$. To find the stationary phase we pick adapted coordinates for the group manifold \cite{Asante:2022lnp}, i.e.\ coordinates $x^I$ for which
\begin{equation}
  \partial_I g= \frac{i}{2}\varsigma_I g\,,\quad \diff g= (4\pi)^{-2}\,\diff x^1\wedge \diff x^2 \wedge \diff x^3\,;
\end{equation}
it then follows that
\begin{align}
  \partial_I^{(b)} \sum_{a< b} S_{ab}=0 \quad \Leftrightarrow \quad  \sum_{a|a \neq  b}&s_{ab} \epsilon_{ab} \Bigg[\frac{\braket{l^+_{ab}|g_a^\dagger \sigma_3 \varsigma_I g_b|l^-_{ba}}}{\braket{l^+_{ab}|g_a^\dagger \sigma_3 g_b|l^-_{ba}}} \nonumber \\
   &-i \braket{l^+_{ab}|g_a^\dagger \sigma_3 g_b|l^+_{ba}} \braket{l^+_{ab}|g_a^\dagger \sigma_3 \varsigma_I g_b|l^+_{ba}} \Bigg]=0\,.
\end{align}
The symbol $\epsilon_{ab}$ above stands for a sign depending on the assumed orientation of the pairings in the amplitude. With the conventions of equation \eqref{amp2}, $\epsilon_{ab}$ is negative whenever $a>b$. Taken together, the stationary and reality conditions may be written in the simple form
\begin{align}
\Re S_{ab}=0 \quad &\Leftrightarrow \quad  g_b\ket{l^+_{ba}}=\vartheta_{ab} g_a\ket{l^+_{ab}}\,, \quad  \label{su11glue} \\
\forall b\,, \; \partial_I^{(b)}  \sum_{a|a\neq b} S_{ab}=0 \quad &\Leftrightarrow  \quad \forall b\,,\; \sum_{a|a\neq b}s_{ab} \epsilon_{ab}\braket{l^+_{ba}|\sigma_3 \varsigma^I |l^-_{ba}}=0\,, \label{su11close}
\end{align}
which the reader may recognize as the customary (see e.g.\ \cite{Barrett:2009mw}) \textit{gluing} and \textit{closure} conditions, respectively. The need for the constraint $\mathcal{C}$ of equation \eqref{constr} should now be clear: since the object $\braket{l^+_{ab}|g_a^\dagger \sigma_3 g_b|l^-_{ba}}$ is always real by virtue of the structure of $\mathrm{SU}(1,1)$ representations, in the absence of $\mathcal{C}$ there would be no real part of the action to maximize, and therefore no gluing equation \eqref{su11glue}. 

Regarding closure \eqref{su11close}, the spin homomorphism 
\begin{equation}
\begin{gathered}
\pi: \mathrm{SU}(1,1) \rightarrow \mathrm{SO}_0(1,2) \\
g \sigma_i g^\dagger = \pi(g)_{ji}\eta^{jk}\sigma_k \,,
\end{gathered}
\end{equation}
with $\eta=\mathrm{diag}(1,-1,-1)$, yields the equivalent equation 
\begin{equation}
\label{closuresu11}
  \forall b\,,\; \sum_{a|a\neq b}s_{ab} \epsilon_{ab} \left[\pi(n_{ba})\hat{e}_2\right] =0 \,. 
\end{equation}
Since $\mathrm{SO}(1,2)$ acts transitively on the one-sheeted space-like hyperboloid $H^{\mathrm{sl}}\ni \hat{e}_2$, the vectors $v_{ba}:=\pi(n_{ba})\hat{e}_2$ are all elements of $H^{\mathrm{sl}}$. Turning to the gluing condition, note first that \eqref{su11glue} also implies $g_b\ket{l^-_{ba}}=\vartheta_{ab}^{-1}g_a\ket{l^-_{ab}}$, from where
\begin{equation}
\label{gluingsu11}
  g_b \ket{l^+_{ba}}\bra{l^-_{ba}}g_b^\dagger = g_a \ket{l^+_{ab}}\bra{l^-_{ab}}g_a^\dagger \quad \Leftrightarrow \quad   \pi(g_b) v_{ba}=\pi(g_a) v_{ab}\,,
\end{equation} 
again employing the spin homomorphism.

Equations \eqref{gluingsu11} and \eqref{closuresu11} afford a geometrical interpretation for the dominant configurations in the vertex amplitude. Minkowski's theorem on convex polyhedra in $\mathbb{R}^{1,2}$  \cite{Simao:2021qno} guarantees that, for all $v_{a\bar b}$ not colinear (fixing $\bar b$), equation \eqref{closuresu11} holds if and only if there exists (up to rigid body motions) a triangle with unit edge vectors $\epsilon_{a \bar b}v_{a \bar b}$ and edge lengths $s_{a \bar b}$. The vertex amplitude is thus suppressed if the boundary data is not in correspondence with four geometrical triangles at the boundary. In its turn the gluing equation \eqref{gluingsu11} dictates that the amplitude is dominated by configurations in which the triangles are $\mathrm{SO}(1,2)$-rotated into coinciding edges. We show in the  appendix that the gluing equations admit two kinds of solutions when the boundary data so allows: one either recovers a degenerate tetrahedron (i.e. a triangle) or a proper tetrahedron (and its reflected counterpart). It is in this sense that one can make the claim that the space-like vertex amplitude induces geometricity in the semiclassical regime.

Finally, applying Hörmander's theorem we obtain an explicit expression for the asymptotic amplitude when the boundary data is that of a tetrahedron with space-like edges. If the Hessian $H$ of the action is non-singular at the two critical points discussed in the appendix, the asymptotic amplitude reads
\begin{align}
\label{asy}
  \scalebox{0.35}{ \input{Pics/vertexsu2_black.tikz}}^{\oplus}_\Lambda= e^{i\frac{\pi}{4}}\frac{\Lambda^{\frac{3}{2}} \gamma^6}{(2\pi)^{\frac{15}{2}}} &\left[\prod_{a< b} s_{ab}\tanh (\pi \Lambda s_{ab})\right]  \nonumber \\
  &\cdot \, \left(\frac{1}{H_{\mathds{1}}^{1/2}} + \frac{e^{\sum_{a\neq b}(-1+2i\Lambda s_{ab}) \theta_{ab}}}{H_{\theta}^{1/2}}\right)+\mathcal{O}\left(\Lambda^{\frac{1}{2}}\right)\,.
\end{align} 
The parameter $\theta_{ab}:=\ln \vartheta_{ab}$ is the Lorentzian dihedral angle \cite{Alexandrov:2000} between the faces $a,b$ of the reconstructed tetrahedron, 
\begin{equation}
  \scalebox{0.7}{ \input{Pics/tetraABCv.tikz}}\,,
\end{equation}
defined such that its sign agrees with equation \eqref{dih} of the appendix. As intended, one finds in equation \eqref{asy} the Regge action $S_R=2\Lambda s_{ab} \theta_{ab}$ with edge lengths $2 \Lambda s_{ab}$ and dihedral angles $\theta_{ab}$.  

Regarding the case of a mixed $\oplus\ominus$ amplitude, it is enough to note that replacing any $\oplus$ pairing by a $\ominus$ pairing in the analysis above leaves equations \eqref{su11glue} and \eqref{su11close} invariant. A mixed space-like amplitude retains the same critical points relative to the $\oplus$-type, and thus the same reconstructed geometry.  The asymptotic expression \eqref{asy}, on the other hand, is modified by interchanging the sign of the relevant angle $\theta^{\oplus} \rightarrow -\theta^{\ominus}$; consequently, for a fixed convention of dihedral angle signs, the mixed semiclassical amplitude fails to reproduce the correct Regge action for the geometric configuration it would otherwise represent. Only $\oplus$-type and $\ominus$-type amplitudes display an appropriate semiclassical behavior.

\section{Discussion}

We have proposed a spin-foam vertex for 3d Lorentzian quantum gravity with both space- and time-like boundaries. The construction hinges on the introduction of two new objects: continuous series coherent states with complex eigenvalues \eqref{r3}, and a Gaussian gluing constraint \eqref{constr}. The Lorentzian Regge action was shown to be recovered in the semiclassical limit. We conclude with a number of final remarks.

\begin{enumerate}

 \item Regarding the gluing constraint $\mathcal{C}$ defined in \eqref{constr}, its inclusion in space-like pairings \eqref{d2} is unjustified in the traditional spin-foam framework: its presence does not follow from a direct manipulation of the partition function for 3-dimensional gravity \eqref{partition}. The discussion of section \ref{sec:asy}, however, shows that its inclusion is paramount for the amplitude to have the right asymptotic behavior. One may wonder whether the need for $\mathcal{C}$ is caused by the particular choice of coherent states which defines the model. However, since the exact same lack of sufficient constraint has been observed in the 4d CH amplitude with time-like faces \cite{Liu:2018gfc, Simao:2021qno}, we find it more likely to rather follow from the structure of the $\mathrm{SU}(1,1)$ continuous series. It remains unknown to us whether a deeper reason for this insufficiency exists, and whether a more natural cure can be found. 
 
\item Concerning the definition of the partition function of equation \eqref{partition}, eventually leading to the proposed amplitude, one may wonder about the fate of the $\mathrm{SU}(1,1)$-gauge and diffeomorphism symmetries of the classical theory. The former causes a divergence of the vertex amplitude due to the non-compactness of the Lie manifold; we have addressed this divergence by gauge-fixing one of the Haar integrals in \eqref{A3dsl} and \eqref{A3dtl}, as prescribed originally by \cite{Engle:2008ev}. On the latter diffeomorphism symmetry, it was shown in \cite{Freidel:2002dw} for $\mathrm{SU}(2)$ $BF$ and the Ponzano-Regge model that the vertex translational symmetry of the classical theory must be modded out from the amplitude, and that this requires dividing by an infinite gauge volume, moreover ensuring topological invariance of the amplitude. We fully expect the same argument to apply to Lorentzian theory, and we leave a detailed analysis of this point for future work. 
 
  \item The expected Regge action \cite{Asante:2021phx} for a space-like tetrahedron $iS=i\sum s_{ab} \theta_{ab}$ figures in the asymptotic formula, but so does an additional imaginary term $i\mathfrak{I}:=-\frac{1}{2}\sum \theta_{ab}$. The presence of $\mathfrak{I}$ is a peculiarity of the model, and it can be traced back to the real part of the complex spin $j=-1/2+is$. There exist at least to ways to interpret this result. The most immediate is to assume the model describes a tetrahedron with complex lengths\footnote{It is perhaps interesting to note that the continuous series Casimir can be written as the square $Q=(x_a p_b \sigma_3^{ab})^2$ of complementary real spinors $\{x^a,p^b\}=i\sigma_3^{ab}$ \cite{Wieland:2020ogk}. The naive square root carries an ordering ambiguity, and its symmetrization $D_a:=(x_ap_a+p_ax_a)/2=x_ap_a-i/2$ is an operator with continuous real spectrum displaced by $i/2$. In this sense the additional $i/2$ term rendering spatial lengths complex may be a consequence of operator ordering ambiguities, and thus a strictly quantum effect. We owe the reviewer this observation.} given by $-ij=s+i/2$; however odd, this would be in agreement with the space-like area spectrum (now understood as  length), which reads $A^2=-(s^2+1/4)$ according to \cite{Freidel:2002hx, Conrady:2010kc}. The imaginary part would be fixed, and small compared to the spins in the regime where gravity is expected to be recovered. The difficulty in assigning a meaningful interpretation to a complex length is an obvious setback. The second possibility is to interpret $\mathfrak{I}$ as part of the amplitude's measure. Since the measure already depends on spins and angles via the Hessian determinant, doing so should simply be a matter of convention. The exact same phenomenon is already present in the time-like CH amplitude \cite{Liu:2018gfc, Simao:2021qno}.

\item The primary purpose of the $(2+1)$ amplitude is to serve as a case study for an eventual complete 4-dimensional Lorentzian theory. To that end one identifies two necessary ingredients from our analysis: the need for some regularization procedure in the time-like amplitude is expected, as is having to enforce additional constraints. The properties of the boundary coherent states here introduced may also justify their inclusion in the 4d theory. 

\item The explicit formulas for both the amplitude and its asymptotics allow for numerically studying configurations involving space- and time-like boundaries, a research direction which has remained unexplored due to the obstacles of the CH time-like amplitude. This opens the door to e.g.\ 1) c omparative studies between the spin-foam framework and that of Causal Dynamical Triangulations \cite{Loll:2019rdj}, potentially bridging the two approaches; and 2) explorations of cosmological scenarios requiring both space- and time-like regions, as is the case for the FRW universe.   

\item Earlier $(2+1)$ state-sums of the Ponzano-Regge type (in that the boundary data consists solely of spin labels) have been proposed\footnote{We thank J. W. Barret for pointing out the relevant literature.}; the associated amplitudes correspond to tetrahedra with entirely time-like \cite{Davids:1998bp, Davids:2000kz} or entirely space-like edges \cite{Garcia-Islas:2003ges}. Our model extends these proposals to mixed edges, and the coherent-state formulation allows for a straightforward generalization to higher-valent polyhedra. 

\end{enumerate} 

{\centering \noindent\rule{2cm}{0.3pt} \\~\\}

{\small \noindent J.D.S. gratefully acknowledges support by the Deutsche Forschungsgemeinschaft (DFG,
German Research Foundation) - Projektnummer/project-number 422809950. The author benefited from useful discussions with A. Jercher and S. Asante on the topic of spin-foam causal ordering.}

\appendix

\section*{Appendix: critical points in the 3d model} 
\addcontentsline{toc}{section}{\protect\numberline{}Appendix: critical points in the 3d model}

\noindent This appendix contains the proof that the parameters $\vartheta_{ab}$ appearing in the asymptotic analysis of the space-like 3-dimensional amplitude are related to the dihedral angles of the reconstructed tetrahedron. We shall take non-colinear boundary data, i.e. data for which Minkowski's theorem is applicable at each triple of boundary states, and we fix $g_4=\mathds{1}$.

Defining $\theta_{ab}:=\ln \vartheta_{ab}$, the gluing condition \eqref{su11close} implies the system of equations
\begin{equation}
  \begin{cases}
  n_{ab}^{-1}g_a^{-1}g_b n_{ba} \ket{l^+}=e^{\theta_{ab}}\ket{l^+} \\
  n_{ab}^{-1}g_a^{-1}g_b n_{ba} \ket{l^-}=e^{-\theta_{ab}}\ket{l^-}\,, \quad n,g\in \mathrm{SU}(1,1)\,,
  \end{cases}
\end{equation}
from where, since $\ket{\pm}$ spans $\mathbb{C}^2$, one sees that $n_{ab}^{-1}g_a^{-1}g_b n_{ba}= e^{\theta_{ab}\sigma_1}$. Then the following chain of equalities holds,
\begin{align}
g_a^{-1}g_b n_{ba} n_{ab}^{-1} &= n_{ab}  e^{\theta_{ab}\sigma_1} \sigma_3 n_{ab}^\dagger \sigma_3 \nonumber\\
&= \cosh \theta_{ab}-i \sinh \theta_{ab}\, n_{ab} \sigma_2 n_{ab}^\dagger \, \sigma_3 \nonumber \\
&=  \cosh \theta_{ab}-i \sinh \theta_{ab}\,  \sigma_3\,  (v_{ab}\cdot \varsigma) \,\sigma_3 \nonumber\\
&= e^{-i \theta_{ab}  \sigma_3\,  (v_{ab}\cdot \varsigma) \,\sigma_3 } \,,
\end{align}
with $v_{ab}=\pi(n_{ab})\hat{e}_2\in H^{\mathrm{sl}}$ the geometrical vector associated to $n_{ab}$. The dot $(\cdot)$ stands for the scalar product with respect to $\eta_{(1,2)}$. We can construct the system of equations
 \begin{equation}
 \label{system2}
  \begin{cases}
  g_a^{-1}g_b n_{ba} n_{ab}^{-1}=e^{-i \theta_{ab}  \sigma_3\,  (v_{ab}\cdot \varsigma)\,\sigma_3 }  \\
  g_c^{-1}g_b n_{bc} n_{cb}^{-1}=e^{-i \theta_{cb}  \sigma_3\,  (v_{cb}\cdot \varsigma)\,\sigma_3 } \\
  g_a^{-1}g_c n_{ca} n_{ac}^{-1}=e^{-i \theta_{ac}  \sigma_3\,  (v_{ac}\cdot \varsigma)\,\sigma_3 } \,,
  \end{cases}
\end{equation}
and by factoring out $g_a,g_b,g_c$ find
\begin{equation}
\label{system1}
  e^{-i \theta_{ab}  \sigma_3\,  (v_{ab}\cdot \varsigma)\,\sigma_3 } n_{ab}n_{ba}^{-1}=e^{-i \theta_{ac}  \sigma_3\,  (v_{ac}\cdot \varsigma)\,\sigma_3 }n_{ac}n_{ca}^{-1}\, e^{-i \theta_{cb}  \sigma_3\,  (v_{cb}\cdot \varsigma)\,\sigma_3 }  n_{cb}n_{bc}^{-1}\,.
\end{equation}
To proceed we make the simplifying assumption that all matched boundary data are parallel, i.e. $n_{ab}=n_{ba}$; there is no loss of generality in doing so since, for a given solution to the gluing equations, each triple of boundary data can be gauge-rotated such that our assumption is satisfied. Equation \eqref{system1} thus implies
\begin{align}
\cosh \theta_{ab}-i\sinh  \theta_{ab} \sigma_3\,  (v_{ab}\cdot \varsigma)\,\sigma_3 =[&\cosh \theta_{ac}-i\sinh  \theta_{ac} \sigma_3\,  (v_{ac}\cdot \varsigma)\,\sigma_3] \nonumber \\ 
&\cdot \, \left[\cosh \theta_{cb}-i\sinh  \theta_{cb} \sigma_3\,  (v_{cb}\cdot \varsigma)\,\sigma_3 \right]\,.
\end{align}
The Pauli matrices together with the identity are linearly independent, so that the previous equation splits into
\begin{equation}
  \begin{cases}
  \cosh \theta_{ab} = \cosh \theta_{ac}\cosh \theta_{cb}-\sinh  \theta_{ac} \sinh  \theta_{cb}\, v_{ac}\cdot v_{cb} \\
  \sinh  \theta_{ab} \,  v_{ab} = \cosh \theta_{ac} \sinh  \theta_{cb} \,  v_{cb} + \cosh \theta_{cb} \sinh  \theta_{ac} \,  v_{ac} - \sinh  \theta_{ac} \sinh  \theta_{cb}\, v_{ac}\times v_{cb}\,,
  \end{cases}
\end{equation}
where the completeness identity
\begin{equation}
  \label{varsigma}
    \varsigma^i \varsigma^j = \eta^{ij} -i \epsilon^{ijk}\eta_{kl} \varsigma^l
  \end{equation}
was used. Contracting the second equation with $v_{ab}\times v_{ac}$ yields 
\begin{equation}
 \theta_{cb}=0 \quad \vee \quad \tanh \theta_{ac}=\frac{v_{cb} \cdot v_{ab}\times v_{ac}}{(v_{ac}\times v_{cb})\cdot(v_{ab}\times v_{ac})}\,,
\end{equation}
or, employing the quadruple product identity $(a\times b)\times (c\times d)=a\cdot(b\times d)c- a\cdot(b\times c)d$, 
\begin{equation}
\label{tetraeqs}
 \theta_{cb}=0 \quad \vee \quad  \tanh \theta_{ac}=\frac{v_{ac}\cdot\left[(v_{cb}\times v_{ac})\times(v_{ac}\times v_{ab})\right]}{(v_{ac}\times v_{cb})\cdot(v_{ab}\times v_{ac})} \,.
\end{equation}

Equations \eqref{tetraeqs} identify two \textit{possible} solutions to the gluing equations. The first is given by $\theta_{cb}=0$, or equivalently $g_b=g_c$. It then follows from the system \eqref{system2} that
\begin{equation}
  e^{-i \theta_{ac}  \sigma_3\,  (v_{ac}\cdot \varsigma)\,\sigma_3 } = e^{-i \theta_{ab}  \sigma_3\,  (v_{ab}\cdot \varsigma)\,\sigma_3 }\,, 
\end{equation}
and - since $v_{ac}$ and $v_{ab}$ must not be colinear, as we assumed that the boundary data was not degenerate - it must be that also\footnote{Curiously, were we working with the time-like model the angles would be Euclidian, and there would be a second solution $\theta_{ab}=\pi$ showing that $g_a=\pm \mathds{1}$. The sign is geometrically irrelevant, since the spin homomorphism is defined modulo $\mathbb{Z}_2$. For hyperbolic functions the solution $g_a=-\mathds{1}$ is absent.} $\theta_{ac}=\theta_{ab}=0$. The same argument can be repeated with a fourth label $d$, from where it follows that $g_a=g_b=g_c=g_d$. By virtue of the gauge fixing $g_4=\mathds{1}$, one finally has that $g_a=\mathds{1}$ for all $a=1,..,4$. Note that this solution is always present whenever the boundary data allows for a non-empty solution set. 

Turning to the second equation in \eqref{tetraeqs}, observe first that the gluing equations \eqref{gluingsu11} essentially describe a triangular development $ \scalebox{0.4}{\input{Pics/dev.tikz}}$ with identified edges and edge orientations. A moment of thought is enough to convince oneself that such a net can either correspond to 1) a degenerate tetrahedron (i.e. a triangle) or 2) to a proper tetrahedron and its reflection (depending on whether the ``flaps'' are closed above or below the bottom face)
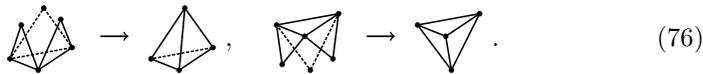
\begin{equation}
  \scalebox{0.6}{ \input{Pics/nets1.tikz}}\,, \quad \scalebox{0.6}{ \input{Pics/nets2.tikz}}\,.
\end{equation}
Constructive inspection shows that options 1) and 2) are mutually exclusive when the boundary data is not made up of four copies of an equilateral triangle: if the edge orientations are such that the net can be closed into a flat triangle, than it cannot correspond to a tetrahedron, and vice-versa. Moreover up to orientation signs (which depend on the particular boundary data) the vectors $v_{ab}$ can be identified with the sides of a (possibly degenerate) tetrahedron,
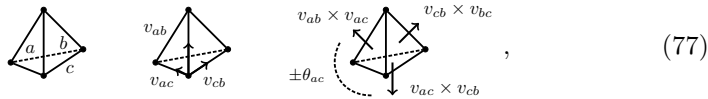
\begin{equation}
  \scalebox{0.7}{ \input{Pics/tetraABCv.tikz}}\,,
\end{equation}
and the second equation of \eqref{tetraeqs} clearly shows that $\theta_{ab}$ labels the dihedral angle between the triangular faces $a$ and $b$ modulo a sign. Thus the following two alternatives are possible: if the boundary data is that of a degenerate tetrahedron then all dihedral angles vanish, and the first and second solutions are simply identified; in that case there is a single solution to the gluing equations given by all $g_{a}=\mathds{1}$. If however the boundary data is that of a proper tetrahedron then its reflected counterpart solves the second equation of \eqref{tetraeqs} with
\begin{equation}
\label{dih}
  g_a=e^{-i \theta_{4a}  \sigma_3\,  (v_{4a}\cdot \varsigma)\,\sigma_3 }\,, \quad \theta_{ac}=\mathrm{arctanh}\, \frac{v_{cb} \cdot v_{ab}\times v_{ac}}{(v_{ac}\times v_{cb})\cdot(v_{ab}\times v_{ac})}
\end{equation}
and this constitutes a different solution from $g_{a}=\mathds{1}$ (which characterizes the original tetrahedron associated to the boundary data). Finally, in the marginal case where the boundary data is that of four copies of the same equilateral triangle, the net can be closed into a flat degenerate tetrahedron, into an equilateral tetrahedron or into its reflection. But since both proper tetrahedra have the exact same dihedral angles their associated critical points coalesce, and there is again a total of two critical configurations.

\bibliographystyle{utphys}
\small 
\bibliography{bib.bib}

\end{document}

%% file: Pics/Style.tikzstyles
\tikzstyle{Node}=[fill=black, draw=black, shape=circle, scale=0.3px]
\tikzstyle{coh}=[fill=white, draw=black, shape=circle, scale=0.6px, line width=1px]
\tikzstyle{coh_big}=[fill=white, draw=black, shape=circle, scale=0.6px]
\tikzstyle{coh_black}=[fill=black, draw=black, shape=circle, scale=0.6px, line width=1px]
\tikzstyle{coh_blue}=[fill=blue, draw=blue, shape=circle, tikzit fill=blue, scale=0.5px]
\tikzstyle{coh_wb}=[fill=white, draw=blue, shape=circle, scale=0.6px, line width=1px]

\tikzstyle{line}=[-, fill=none, line width=1px]
\tikzstyle{blockline}=[-, fill=black, line width=3px]
\tikzstyle{boost}=[-, fill={rgb,255: red,128; green,128; blue,128}, draw={rgb,255: red,128; green,128; blue,128}, tikzit fill={rgb,255: red,128; green,128; blue,128}, tikzit draw={rgb,255: red,128; green,128; blue,128}, line width=5px]
\tikzstyle{specialsu2}=[-, line width=5px, fill=black, draw={rgb,255: red,0; green,0; blue,189}, tikzit fill=white, tikzit draw=black]
\tikzstyle{boxdash}=[-, line width=5px, fill=black, dash pattern=on 2pt off 2pt]
\tikzstyle{arrow}=[line width=1.1px, ->]
\tikzstyle{arrowdotted}=[line width=1.1px, ->, dash pattern=on 2pt off 1.5pt]
\tikzstyle{dashed}=[-, line width=1px, dash pattern=on 2pt off 1pt, fill=none]
\tikzstyle{thindash}=[-, line width=0.5px, dash pattern=on 1pt off 3pt, draw={rgb,255: red,158; green,158; blue,158}]
\tikzstyle{grayfill}=[-, fill={rgb,255: red,234; green,234; blue,234}, draw=none]
\tikzstyle{thingray}=[-, line width=0.8px, draw={rgb,255: red,158; green,158; blue,158}]
\tikzstyle{arrowgray}=[draw={rgb,255: red,158; green,158; blue,158}, ->, line width=1px]
\tikzstyle{line_blue}=[-, line width=1.5px, draw=blue, tikzit draw=blue]
\tikzstyle{blue_dashed}=[-, draw=blue, tikzit draw=blue, line width=1.5px, dash pattern=on 2pt off 1.5pt]

%% file: Pics/cstate.tikz
\begin{tikzpicture}
	\begin{pgfonlayer}{nodelayer}
		\node [style=none] (0) at (-0.5, 0.25) {};
		\node [style=coh] (2) at (0.5, 0.25) {};
	\end{pgfonlayer}
	\begin{pgfonlayer}{edgelayer}
		\draw [style=line] (0.center) to (2);
	\end{pgfonlayer}
\end{tikzpicture}

%% file: Pics/vertexsu2.tikz
\begin{tikzpicture}
	\begin{pgfonlayer}{nodelayer}
		\node [style=coh] (1) at (-0.75, 2.75) {};
		\node [style=coh] (2) at (0, 2.75) {};
		\node [style=coh] (3) at (0.75, 2.75) {};
		\node [style=none] (20) at (-1.25, 2) {};
		\node [style=none] (21) at (1.25, 2) {};
		\node [style=coh] (60) at (-0.75, -2.25) {};
		\node [style=coh] (61) at (0, -2.25) {};
		\node [style=coh] (62) at (0.75, -2.25) {};
		\node [style=coh] (63) at (-2.5, -0.5) {};
		\node [style=coh] (64) at (-2.5, 0.25) {};
		\node [style=coh] (65) at (-2.5, 1) {};
		\node [style=coh] (66) at (2.5, -0.5) {};
		\node [style=coh] (67) at (2.5, 0.25) {};
		\node [style=coh] (68) at (2.5, 1) {};
		\node [style=none] (69) at (-1.25, -1.5) {};
		\node [style=none] (70) at (1.25, -1.5) {};
		\node [style=none] (71) at (1.75, 1.5) {};
		\node [style=none] (72) at (1.75, -1) {};
		\node [style=none] (73) at (-1.75, 1.5) {};
		\node [style=none] (74) at (-1.75, -1) {};
	\end{pgfonlayer}
	\begin{pgfonlayer}{edgelayer}
		\draw [style=blockline] (20.center) to (21.center);
		\draw [style=blockline] (21.center) to (20.center);
		\draw [style=blockline] (69.center) to (70.center);
		\draw [style=blockline] (70.center) to (69.center);
		\draw [style=blockline] (72.center) to (71.center);
		\draw [style=blockline] (74.center) to (73.center);
		\draw [style=line] (2) to (61);
		\draw [style=line] (64) to (67);
		\draw [style=line, bend right=45, looseness=1.75] (3) to (68);
		\draw [style=line, bend left=45, looseness=1.75] (1) to (65);
		\draw [style=line, bend left=45, looseness=1.75] (63) to (60);
		\draw [style=line, bend right=45, looseness=1.75] (66) to (62);
	\end{pgfonlayer}
\end{tikzpicture}

%% file: Pics/braket.tikz
\begin{tikzpicture}
	\begin{pgfonlayer}{nodelayer}
		\node [style=coh] (1) at (-0.75, 0.25) {};
		\node [style=coh] (5) at (0.75, 0.25) {};
		\node [style=none] (6) at (0, 0.5) {};
		\node [style=none] (7) at (0, 0) {};
	\end{pgfonlayer}
	\begin{pgfonlayer}{edgelayer}
		\draw [style=line] (1) to (5);
		\draw [style=blockline] (6.center) to (7.center);
	\end{pgfonlayer}
\end{tikzpicture}

%% file: Pics/braket_black1.tikz
\begin{tikzpicture}
	\begin{pgfonlayer}{nodelayer}
		\node [style={coh_black}] (1) at (-0.75, 0.25) {};
		\node [style=coh] (5) at (0.75, 0.25) {};
		\node [style=none] (6) at (0, 0.5) {};
		\node [style=none] (7) at (0, 0) {};
	\end{pgfonlayer}
	\begin{pgfonlayer}{edgelayer}
		\draw [style=line] (1) to (5);
		\draw [style=blockline] (6.center) to (7.center);
	\end{pgfonlayer}
\end{tikzpicture}

%% file: Pics/braket_black2.tikz
\begin{tikzpicture}
	\begin{pgfonlayer}{nodelayer}
		\node [style=coh] (1) at (-0.75, 0.25) {};
		\node [style={coh_black}] (5) at (0.75, 0.25) {};
		\node [style=none] (6) at (0, 0.5) {};
		\node [style=none] (7) at (0, 0) {};
	\end{pgfonlayer}
	\begin{pgfonlayer}{edgelayer}
		\draw [style=line] (1) to (5);
		\draw [style=blockline] (6.center) to (7.center);
	\end{pgfonlayer}
\end{tikzpicture}

%% file: Pics/vertexsu2_black.tikz
\begin{tikzpicture}
	\begin{pgfonlayer}{nodelayer}
		\node [style=coh] (1) at (-0.75, 2.75) {};
		\node [style=coh] (2) at (0, 2.75) {};
		\node [style={coh_black}] (3) at (0.75, 2.75) {};
		\node [style=none] (20) at (-1.25, 2) {};
		\node [style=none] (21) at (1.25, 2) {};
		\node [style=coh] (60) at (-0.75, -2.25) {};
		\node [style={coh_black}] (61) at (0, -2.25) {};
		\node [style=coh] (62) at (0.75, -2.25) {};
		\node [style={coh_black}] (63) at (-2.5, -0.5) {};
		\node [style={coh_black}] (64) at (-2.5, 0.25) {};
		\node [style={coh_black}] (65) at (-2.5, 1) {};
		\node [style={coh_black}] (66) at (2.5, -0.5) {};
		\node [style=coh] (67) at (2.5, 0.25) {};
		\node [style=coh] (68) at (2.5, 1) {};
		\node [style=none] (69) at (-1.25, -1.5) {};
		\node [style=none] (70) at (1.25, -1.5) {};
		\node [style=none] (71) at (1.75, 1.5) {};
		\node [style=none] (72) at (1.75, -1) {};
		\node [style=none] (73) at (-1.75, 1.5) {};
		\node [style=none] (74) at (-1.75, -1) {};
	\end{pgfonlayer}
	\begin{pgfonlayer}{edgelayer}
		\draw [style=blockline] (20.center) to (21.center);
		\draw [style=blockline] (21.center) to (20.center);
		\draw [style=blockline] (69.center) to (70.center);
		\draw [style=blockline] (70.center) to (69.center);
		\draw [style=blockline] (72.center) to (71.center);
		\draw [style=blockline] (74.center) to (73.center);
		\draw [style=line] (2) to (61);
		\draw [style=line] (64) to (67);
		\draw [style=line, bend right=45, looseness=1.75] (3) to (68);
		\draw [style=line, bend left=45, looseness=1.75] (1) to (65);
		\draw [style=line, bend left=45, looseness=1.75] (63) to (60);
		\draw [style=line, bend right=45, looseness=1.75] (66) to (62);
	\end{pgfonlayer}
\end{tikzpicture}

%% file: Pics/tetraABCv.tikz
\begin{tikzpicture}
	\begin{pgfonlayer}{nodelayer}
		\node [style=Node] (0) at (1.5, 0) {};
		\node [style=Node] (1) at (-1.25, -0.5) {};
		\node [style=Node] (2) at (0, 1.5) {};
		\node [style=Node] (3) at (0, -1) {};
		\node [style=none] (4) at (-1, -1.25) {$v_{ac}$};
		\node [style=none] (5) at (-1.25, 0.75) {$v_{ab}$};
		\node [style=none] (6) at (1, -1.25) {$v_{cb}$};
		\node [style=none] (7) at (-0.5, -0.75) {};
		\node [style=none] (8) at (0.75, -0.5) {};
		\node [style=none] (9) at (0, 0.25) {};
		\node [style=Node] (10) at (9, 0) {};
		\node [style=Node] (11) at (6.25, -0.5) {};
		\node [style=Node] (12) at (7.5, 1.5) {};
		\node [style=Node] (13) at (7.5, -1) {};
		\node [style=none] (14) at (6.25, 0.75) {};
		\node [style=none] (15) at (8.75, 1) {};
		\node [style=none] (16) at (8, 0.25) {};
		\node [style=none] (17) at (7, 0) {};
		\node [style=none] (18) at (5.5, 1.25) {$v_{ab}\times v_{ac}$};
		\node [style=none] (19) at (10.25, 1.5) {$v_{cb}\times v_{bc}$};
		\node [style=none] (20) at (7.75, -0.5) {};
		\node [style=none] (21) at (7.75, -1.75) {};
		\node [style=none] (22) at (9.75, -1.5) {$v_{ac}\times v_{cb}$};
		\node [style=none] (23) at (5.75, 0.25) {};
		\node [style=none] (24) at (7, -1.75) {};
		\node [style=none] (25) at (4.5, -1) {\small $\pm \theta_{ac}$};
		\node [style=Node] (26) at (-4, 0) {};
		\node [style=Node] (27) at (-6.75, -0.5) {};
		\node [style=Node] (28) at (-5.5, 1.5) {};
		\node [style=Node] (29) at (-5.5, -1) {};
		\node [style=none] (30) at (-6, 0) {$a$};
		\node [style=none] (31) at (-4.75, 0.25) {$b$};
		\node [style=none] (32) at (-4.5, -0.75) {$c$};
	\end{pgfonlayer}
	\begin{pgfonlayer}{edgelayer}
		\draw [style=line] (2) to (0);
		\draw [style=line] (2) to (1);
		\draw [style=line] (3) to (0);
		\draw [style=line] (2) to (3);
		\draw [style=dashed] (1) to (0);
		\draw [style=line] (3) to (1);
		\draw [style=arrow] (3) to (8.center);
		\draw [style=arrow] (3) to (7.center);
		\draw [style=arrow] (3) to (9.center);
		\draw [style=line] (12) to (10);
		\draw [style=line] (12) to (11);
		\draw [style=line] (13) to (10);
		\draw [style=line] (12) to (13);
		\draw [style=dashed] (11) to (10);
		\draw [style=line] (13) to (11);
		\draw [style=arrow] (17.center) to (14.center);
		\draw [style=arrow] (16.center) to (15.center);
		\draw [style=arrow] (20.center) to (21.center);
		\draw [style=dashed, bend right=60, looseness=1.25] (23.center) to (24.center);
		\draw [style=line] (28) to (26);
		\draw [style=line] (28) to (27);
		\draw [style=line] (29) to (26);
		\draw [style=line] (28) to (29);
		\draw [style=dashed] (27) to (26);
		\draw [style=line] (29) to (27);
	\end{pgfonlayer}
\end{tikzpicture}

%% file: Pics/dev.tikz
\begin{tikzpicture}
	\begin{pgfonlayer}{nodelayer}
		\node [style=Node] (23) at (0.5, -0.5) {};
		\node [style=Node] (24) at (-1.5, -0.5) {};
		\node [style=Node] (26) at (-0.5, 1.25) {};
		\node [style=Node] (27) at (1.5, 1.25) {};
		\node [style=Node] (28) at (2.5, -0.5) {};
		\node [style=Node] (29) at (3.5, 1.25) {};
		\node [style=none] (30) at (-0.5, -0.5) {};
		\node [style=none] (31) at (1.5, -0.5) {};
		\node [style=none] (32) at (2.5, 1.25) {};
		\node [style=none] (33) at (0.5, 1.25) {};
		\node [style=none] (34) at (-1, 0.25) {};
		\node [style=none] (35) at (3, 0.25) {};
	\end{pgfonlayer}
	\begin{pgfonlayer}{edgelayer}
		\draw [style=line] (24) to (23);
		\draw [style=line] (26) to (23);
		\draw [style=line] (24) to (26);
		\draw [style=line] (28) to (23);
		\draw [style=line] (28) to (27);
		\draw [style=line] (27) to (23);
		\draw [style=line] (26) to (27);
		\draw [style=line] (27) to (29);
		\draw [style=line] (29) to (28);
		\draw [style=arrowdotted] (23) to (30.center);
		\draw [style=arrowdotted] (23) to (31.center);
		\draw [style=arrowdotted] (29) to (35.center);
		\draw [style=arrowdotted] (29) to (32.center);
		\draw [style=arrowdotted] (27) to (33.center);
		\draw [style=arrowdotted] (26) to (34.center);
		\draw [style=arrow] (29) to (35.center);
		\draw [style=arrow] (29) to (32.center);
		\draw [style=arrow] (27) to (33.center);
		\draw [style=arrow] (26) to (34.center);
		\draw [style=arrow] (23) to (30.center);
		\draw [style=arrow] (23) to (31.center);
	\end{pgfonlayer}
\end{tikzpicture}

%% file: Pics/nets1.tikz
\begin{tikzpicture}
	\begin{pgfonlayer}{nodelayer}
		\node [style=Node] (10) at (-2, -0.25) {};
		\node [style=Node] (11) at (-4.75, -0.75) {};
		\node [style=Node] (13) at (-3.5, -1.25) {};
		\node [style=Node] (14) at (-2.5, 1) {};
		\node [style=Node] (15) at (-4.25, 0.75) {};
		\node [style=Node] (16) at (-3.25, 1.5) {};
		\node [style=Node] (23) at (4.25, -0.25) {};
		\node [style=Node] (24) at (1.5, -0.75) {};
		\node [style=Node] (25) at (2.75, -1.25) {};
		\node [style=Node] (26) at (3, 1.5) {};
		\node [style=none] (31) at (-0.75, 0.25) {};
		\node [style=none] (32) at (0.5, 0.25) {};
	\end{pgfonlayer}
	\begin{pgfonlayer}{edgelayer}
		\draw [style=line] (13) to (10);
		\draw [style=dashed] (11) to (10);
		\draw [style=line] (13) to (11);
		\draw [style=line] (15) to (11);
		\draw [style=line] (15) to (13);
		\draw [style=line] (14) to (13);
		\draw [style=line] (14) to (10);
		\draw [style=dashed] (10) to (16);
		\draw [style=dashed] (16) to (11);
		\draw [style=line] (25) to (23);
		\draw [style=dashed] (24) to (23);
		\draw [style=line] (25) to (24);
		\draw [style=line] (26) to (25);
		\draw [style=line] (26) to (23);
		\draw [style=line] (24) to (26);
		\draw [style=arrow] (31.center) to (32.center);
	\end{pgfonlayer}
\end{tikzpicture}

%% file: Pics/nets2.tikz
\begin{tikzpicture}
	\begin{pgfonlayer}{nodelayer}
		\node [style=Node] (17) at (-2, 1.25) {};
		\node [style=Node] (18) at (-4.75, 0.75) {};
		\node [style=Node] (19) at (-3.5, 0.25) {};
		\node [style=Node] (20) at (-2.25, -0.75) {};
		\node [style=Node] (21) at (-4.5, -1) {};
		\node [style=Node] (22) at (-3.25, -1.25) {};
		\node [style=Node] (27) at (4.25, 1.25) {};
		\node [style=Node] (28) at (1.5, 0.75) {};
		\node [style=Node] (29) at (2.75, 0.25) {};
		\node [style=Node] (30) at (3, -1.25) {};
		\node [style=none] (33) at (-0.75, 0.25) {};
		\node [style=none] (34) at (0.5, 0.25) {};
	\end{pgfonlayer}
	\begin{pgfonlayer}{edgelayer}
		\draw [style=line] (19) to (17);
		\draw [style=line] (18) to (17);
		\draw [style=line] (19) to (18);
		\draw [style=line] (21) to (18);
		\draw [style=line] (21) to (19);
		\draw [style=line] (20) to (19);
		\draw [style=line] (20) to (17);
		\draw [style=dashed] (17) to (22);
		\draw [style=dashed] (22) to (18);
		\draw [style=line] (29) to (27);
		\draw [style=line] (28) to (27);
		\draw [style=line] (29) to (28);
		\draw [style=line] (30) to (29);
		\draw [style=line] (30) to (27);
		\draw [style=line] (28) to (30);
		\draw [style=arrow] (33.center) to (34.center);
	\end{pgfonlayer}
\end{tikzpicture}